\documentclass{ieeeaccess}
\usepackage{cite}
\usepackage{amsmath,amssymb,amsfonts}
\usepackage{algorithmic}
\usepackage{graphicx}
\usepackage{textcomp}

\usepackage{multirow}
\usepackage{url}

\def\BibTeX{{\rm B\kern-.05em{\sc i\kern-.025em b}\kern-.08em
    T\kern-.1667em\lower.7ex\hbox{E}\kern-.125emX}}
\begin{document}
\history{Date of publication xxxx 00, 0000, date of current version xxxx 00, 0000.}
\doi{10.1109/ACCESS.2022.DOI}

\title{Profiling Television Watching Behaviour Using Bayesian Hierarchical Joint Models for Time-to-Event and Count Data}
\author{\uppercase{Rafael A. Moral}\authorrefmark{1}, \uppercase{Zhi Chen}\authorrefmark{3}, \uppercase{Shuai Zhang}\authorrefmark{3}, \uppercase{Sally McClean}\authorrefmark{3}, \uppercase{Gabriel R. Palma}\authorrefmark{1}, \uppercase{Brahim Allan}\authorrefmark{2} and \uppercase{Ian Kegel}\authorrefmark{2}}
\address[1]{Maynooth University, Maynooth, Ireland}
\address[2]{British Telecommunications, England}
\address[3]{Ulster University, Belfast, Northern Ireland}
\tfootnote{This research is supported by BTIIC (the BT Ireland Innovation Centre), funded by BT and Invest Northern Ireland. We are also thankful to the YouView team for providing the sample data and domain knowledge.}

\markboth
{Moral \headeretal: Profiling Television Watching Behaviour Using Bayesian Models}
{Moral \headeretal: Profiling Television Watching Behaviour Using Bayesian Models}

\corresp{Corresponding author: Rafael A. Moral (e-mail: rafael.deandrademoral@mu.ie).}
.

\begin{abstract}
Customer churn prediction is a valuable task in many industries. In telecommunications it presents great challenges, given the high dimensionality of the data, and how difficult it is to identify underlying frustration signatures, which may represent an important driver regarding future churn behaviour. Here, we propose a novel Bayesian hierarchical joint model that is able to characterise customer profiles based on how many events take place within different television watching journeys, and how long it takes between events. The model drastically reduces the dimensionality of the data from thousands of observations per customer to 11 customer-level parameter estimates and random effects. We test our methodology using data from 40 BT customers (20 active and 20 who eventually cancelled their subscription) whose TV watching behaviours were recorded from October to December 2019, totalling approximately half a million observations. Employing different machine learning techniques using the parameter estimates and random effects from the Bayesian hierarchical model as features yielded up to 92\% accuracy predicting churn, associated with 100\% true positive rates and false positive rates as low as 14\% on a validation set. Our proposed methodology represents an efficient way of reducing the dimensionality of the data, while at the same time maintaining high descriptive and predictive capabilities. We provide code to implement the Bayesian model at \url{https://github.com/rafamoral/profiling_tv_watching_behaviour}.
\end{abstract}


\titlepgskip=-15pt

\maketitle

\section{Introduction}
\label{sec:introduction}
\PARstart{T}{his} paper presents the development and testing of a Bayesian hierarchical model-based approach to characterise the viewing behaviours of BT TV customers. A hierarchical model is appropriate here as each customer has multiple viewing sessions which can be quite diverse both in content and duration. Features extracted from this customer behaviour model can then be used to profile customers in terms of their preferences and engagement. Features extracted from such profiles are then used to classify customers as either frustrated or satisfied, where we hypothesise that customer frustration is often an early indicator of later churn which could be driven by a perceived lack of content choice. The approach is validated using labelled data from customers who subsequently cancelled their subscription. User profiling is a key element in understanding user behaviours and relating these behaviours to subsequent actions. 

BT TV is a subscription IPTV service provided by the BT Group, which uses the YouView television platform, to provide a range of channels and services. An anonymised dataset, based on YouView client logs, was prepared and shared by the YouView Data and Insights team for behavioural data analysis and modelling purposes, as described in this paper. As the aim here is to develop an initial model, the study was conducted on a subset of the data consisting of 20 active customers and 20 customers that subsequently cancelled the service. Overall, there were 32,556 journeys in the sample data from these customers that cancelled. The proposed model and associated results can then serve as a basis of transferring the knowledge to a much bigger dataset from BT customers.

The Youview data in this study are primarily time-stamped system logs concerning sequences of TV customer events of interest. A major focus in BT is to better understand how customer navigate their content and, from that, infer whether these customers are satisfied with the service and identify where they may be frustrated by their engagement. The analysis and modelling of such time-stamped event data has become increasingly popular over recent years, especially for business data where such work falls under the Process Mining umbrella \cite{vanderaalst}. Here, process mining can be defined as a data driven analytics approach used to extract knowledge and insights from the event logs, and use them to discover, monitor and improve the process \cite{sakr}. In our case we use TV customer behaviour data provided by YouView in the form of records that capture the customer interactions with TV from the point when the TV is turned on, to the time when the TV is turned off, including all steps in the path to reach a particular channel; failures, reported as error messages, and changes of channel, are also recorded. This is a rich process, generating multiple and diverse possible trajectories for huge numbers of customers and, as such, is well suited to a mining and modelling approach. 

More specifically, in the current research we use the Youview data to develop and implement a Bayesian hierarchical joint model for time-between-events and number of customer events. The model is implemented so as to profile typical journeys for each customer. It is based on covariates, such as number of viewing events and programme genre. We then use a combination of statistical machine learning methods to classify the customers into ``frustration classes'', where the Bayesian hierarchical joint model parameters and associated covariates serve as additional features for the classification. The customers were subsequently labelled as either ``active'' or ``cancelled'', regarding their subscription. While there is an extensive literature on predicting churn \cite{tanguyanmoro}, particularly for telecoms, our approach is novel in that we provide a hybrid model to proactively predict subscription cancellation. The hierarchical Bayesian model proposed here handles heterogeneous and complex data efficiently. It is used to identify and extract concise latent customer-specific frustration signatures, which are then used for efficient churn classification, thus paving the way for pre-emptive intervention as a strategy to improve customer retention.

\section{Previous Work}
\label{sec:previouswork}
Customer Relationship Management (CRM) is an important aspect of business practice and focuses on developing a loyal and long-term customer base. An important aspect of CRM is customer retention which is often much more difficult and expensive than acquiring new customers. The burgeoning accumulation of massive business data collections, particularly in the telecoms industry, and increasingly mature data analytics technologies have served to accelerate the development and deployment of customer churn models to predict customer behaviour. In a recent survey, \cite{tanguyanmoro} demonstrate that data analytics techniques are becoming increasingly used for customer churn prediction, particular in the telecoms industry, where neural networks, decision trees, support vector machines and logistic regression have been most commonly used as the classification algorithms. \cite{decaigny} also observed that decision trees and logistic regression have both been used frequently in customer churn prediction. However, these algorithms fail to cope well with data complexities and interactions so, as a result, \cite{decaigny} improved on such approaches by introducing the logit leaf model (LLM) which segments the data, before prediction, thus better characterising the feature domain, while retaining previous advantages of strong predictive performance and good comprehensibility.

Since modern-day customers exhibit diverse and highly dynamic behaviours, retention is often highly important to businesses. In such dynamic situations, churn can be caused by customer dissatisfaction, rival products or services, regulations or the customer's personal circumstances which the business cannot influence. The importance of dynamic behaviour-based churn prediction for telecoms was emphasised by \cite{alboukaey} due to both the dynamic nature of the customer activities and their subsequent behaviours. \cite{viol} have recently highlighted the improving opportunities for real-time monitoring and resulting ability to discover changes in psychological patterns. They particularly focus on Change Point Analysis (CPA) and describe a Pattern Transition Detection Algorithm for psychological time series data with pattern transitions. In particular, characterising human behaviour from online data has become increasingly feasible, particularly for Cybersecurity \cite{amato}, where it is important to distinguish between normal and pathological user behaviours, in a timely manner. In our problem domain, IPTV customer logs capture comprehensive customer interactions and activities, facilitating the development of customer behaviour modelling. Such models may then be used to predict future behaviours of interests, such as churn. In such an approach, we can analyse the history of interactions and activities of dissatisfied customers who subsequently churn and compare this with the corresponding profile of a satisfied customer. For example, \cite{borbora} analysed the activity data for Massively Multiplayer Online Role-Playing Games (MMORPGs) to identify three factors of engagement, enthusiasm and persistence which proved useful features for churn prediction; these resonate well with our current focus on IPTV where the customer logs can track engagement with the service in terms of session viewing, enthusiasm as evidenced by volume of activity and persistence in terms of sustained viewing. 
For the telecoms and related online-based industries, the product is a subscription service \cite{kavitha}, where prediction of subscription has been, and still is, a very active research area; in the current paper, churn presents as cancellation of the subscription, as in, for example, \cite{altinisik}. For such churn prediction, with advances in technology, (near) real-time data are becoming increasingly available and can serve to provide early indications of churn and learn when customers have an increased probability of churning \cite{vafeiadis}. The early detection of potential churners facilitates focussing on such customers for targeted interventions \cite{burez}. Customer churn management can be done in either reactive or proactive mode \cite{lalwani}. In the reactive mode, churn occurs first and the customer is then offered an incentive to return. In proactive mode, the probability of churn is predicted from available data, and, if churn is likely, incentives can be offered to increase the likelihood of retention. With the increasing availability of online, (near) real-time data, there are increasing possibilities for using dynamic features to characterise user behaviours and provide improved opportunities for proactive churn prediction.

The desirability of using hybrid/hierarchical approaches to classifying complex churn data from online domains has been highlighted by \cite{mohammadi} where clustering is used to identify churners, and subsequently, a survival model is used to build relevant features into the classification model. \cite{huang} and \cite{decaigny}, inter alia, provide further discussions of the advantage of such hybrid approaches for telecoms churn prediction. 

\section{Methods}

\subsection{Data Preparation}

YouView Data and Insights team initially provided an anonymized dataset containing client event logs from BT YouView set-top boxes (STB) to aid our research. The data contains information about TV customer behaviour in terms of interactive activities between customers and the Electronic Programme Guide (EPG) system on the STB, such as powering on/off the box, tuning into IP channels, beginning/ending playback of Video-on-Demand content, etc. In addition to STB performance information, the data contains error/failure logs. Consequently, the raw data contains multiple streams of events, and we must extract the key data flow that illustrates customer journey paths. In this study, a customer journey refers to the steps a customer takes to navigate the EPG system from the time the STB is turned on until the customer reaches IP channel content.

Some columns in the raw dataset contain crucial information, such as ``EVENT DT TM" displaying timestamps and ``EVENT ID" displaying the event log category for each record. And for other columns such as ``EVENT SPEC 1", include information regarding subcategories for specific main event categories. Using the domain knowledge shared by the YouView Data and Insights team, we create two separate reference tables for the original event logs in order to extract target data streams. Specifically, one table displays a list of unique event IDs for all events generated by customers' ``click" actions on the remote control. In addition to the event ID, the second table contains additional information for each target event log, including the IPTV system's URL address, which indicates the UI context information. Using the two reference tables, we can extract a subset of the raw data comprising clickstream and UI context data. The target event log is then renamed by appending a shorter tag referencing the two tables. The general procedure is depicted in Figure~\ref{fig:dataprep}(a).

\Figure[t!](topskip=0pt, botskip=0pt, midskip=0pt)[width = .9\textwidth]{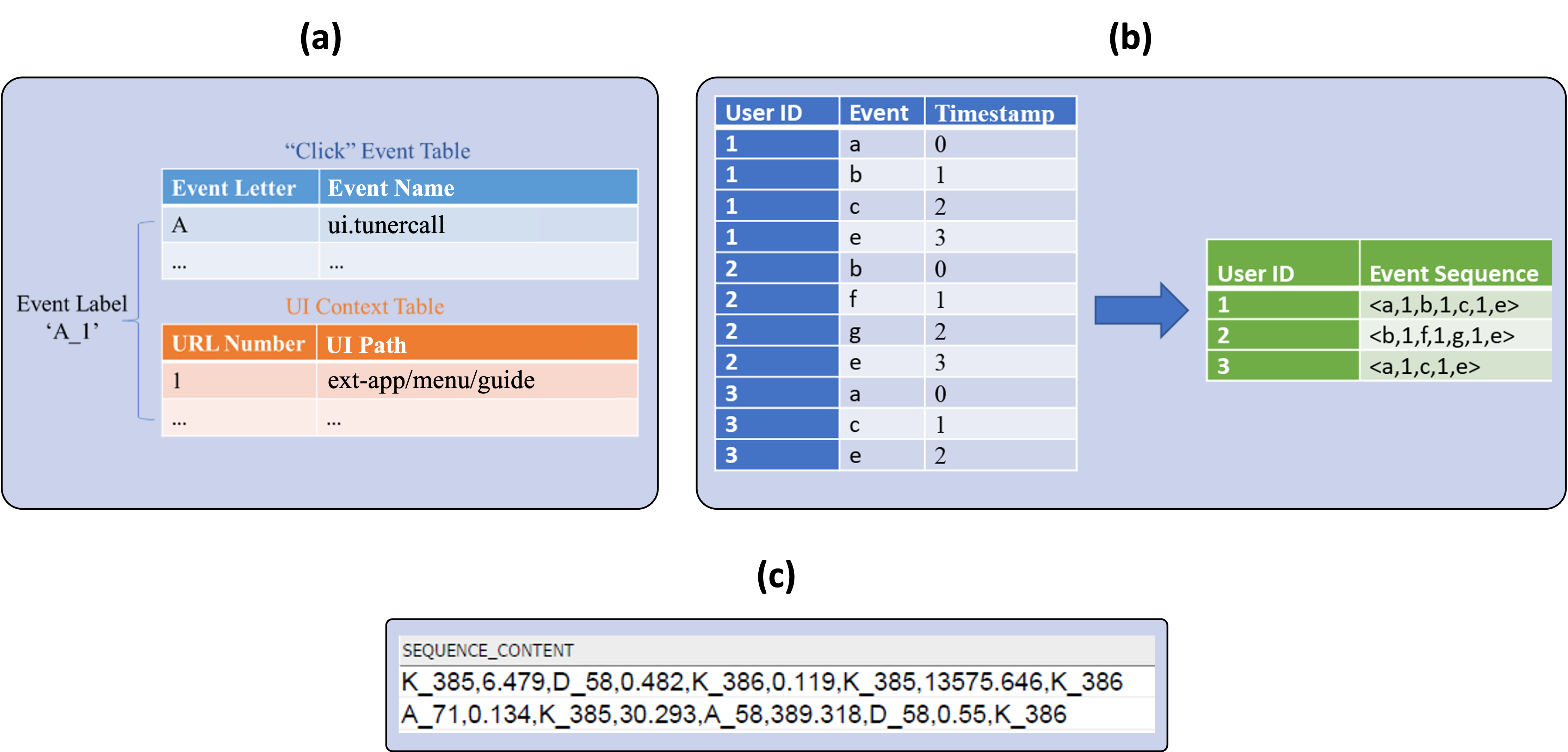}
{Illustration of how the YouView set-top box data was prepared for analysis. We illustrate event-labelling in (a), where the label `A\_1' is created by cross-referencing two tables. In (b), we provide an example of sequence generation from data on user IDs and timestamped events. Finally, in (c) a snapshot of generated journeys/sequences is displayed: in the first row, the customer starts with event `K\_385', then 6.479 seconds later, event `D\_58' is triggered, and so on.\label{fig:dataprep}}

Using the timestamp information from the raw dataset in conjunction with the newly generated labels, we can convert the initial event logs into event sequences for each customer journey on the STB. Each event sequence consists of a series of ``click" events separated by intervals in the format ``event-interval-event." Using dummy data, Figure~\ref{fig:dataprep}(b) provides an illustration of the procedure, while a snapshot of the sequences generated by the real data is shown in Figure~\ref{fig:dataprep}(c).

As a result, we have reformatted the data by generating a new table of event sequences in accordance with the preceding step. In addition, each sequence's start and end times, total duration, channel name, and genre name are included in the new table.

\subsection{The Model}

The hierarchical model was developed with the objective of obtaining customer signatures, or profiles, based on their behaviour and then clustering them into similar groups. By comparing the observed groupings with the known cancellation variable, this can generate insight about a latent frustration signature.

Since journeys are nested within customers, and their features are likely to be correlated, we propose a hierarchical model. We attempt to jointly model the total number of events in a journey and the time between two consecutive events within the journey (see Figure \ref{fig:diagram}). We assume that the time between two events depends on the previous time between two events, i.e. a first-order Markov assumption.

\Figure[t!](topskip=0pt, botskip=0pt, midskip=0pt)[width = .9\textwidth]{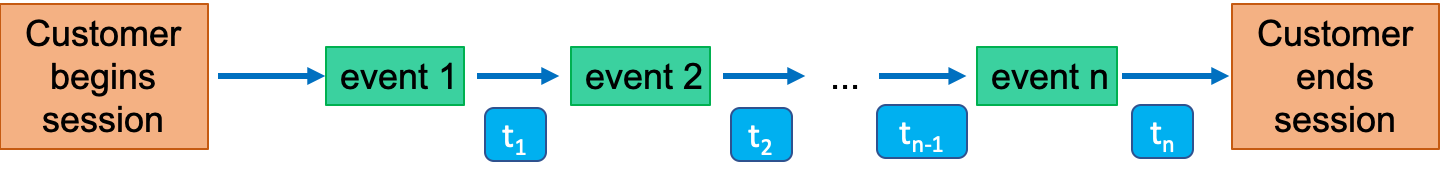}
{A diagram representing one customer journey (session) with $n$ events, where $t$ represents the time between two consecutive events.\label{fig:diagram}}

We assume that the total time per session is a random truncated Poisson sum of gamma random variables. In our notation, $c$ is the index for customer, $c=1,\ldots,C$, $s$ is the index for journey within customer, $s=1,\ldots,S_c$, $i$ is the index for event within journey within customer, $i=1,\ldots,I_{sc}$, $T_{sc}^*$ is the random variable representing the total time per journey, $T_{isc}$ is the random variable representing the duration of an intermediate state (event), and $N_{sc}$ is the number of events (states) per journey. We have that $$T_{sc}^*=\displaystyle\sum_{i=1}^{N_{sc}}T_{isc}.$$

We may write the model for the number of events in a journey as $$N_{sc}|\mathbf{d}_c \sim \mbox{Poisson}^+(\lambda_{sc})$$ where the linear predictor depends on customer-level random effects $\mathbf{d}_c$, and may be written as $$\log(\lambda_{sc} )=d_{0c}+d_{1c}\times(\mbox{predictor at journey level})_{sc},$$ with $\mathbf{d}_c\sim\mbox{N}(\boldsymbol{\delta},\boldsymbol{\Sigma}_d)$ and $\boldsymbol{\Sigma}_d=\mbox{diag}\{\sigma^2_{d_0},\sigma^2_{d_1},\ldots\}$ a diagonal variance-covariance matrix. The time in between events depends on the previous time, and is assumed to follow a gamma distribution, i.e. $$T_{isc} |T_{i-1,sc},\mathbf{b}_c,p_c \sim \mbox{Gamma}(\mu_{isc},\psi_c)$$ with the linear predictor for the mean $\mu_{isc}$ written as the sum of two components: $$\log(\mu_{isc})=\eta_{isc}+\omega_{isc}.$$ The first, $\eta_{isc}$, includes the effects of covariates at the event level, modelled as random effects within customers, and may be written as $$\eta_{isc}=b_{0c}+b_{1c}\times(\mbox{predictor at event level})_{isc},$$ with $\mathbf{b}_c \sim \mbox{N}(\boldsymbol{\beta},\boldsymbol{\Sigma}_b)$, $\boldsymbol{\Sigma}_b=\mbox{diag}\{\sigma^2_{b_0},\sigma^2_{b_1},\ldots\}$. The second component, $\omega_{isc}$, is the autoregressive component, and models the dependence over time with the autoregressive parameter $p_c$, also a customer-level random effect, such that $$\omega_{isc}=p_{sc}\times T_{i-1,sc},$$ with $p_{sc}\sim\mbox{N}(\phi_c,\sigma_p^2)$.

\subsection{Case Study}

The original data provided by YouView contains approximately 100 million event logs from more than 10,000 customers over a three-month period (October to December 2019), which is large and complex data. The data also includes labelled customers, i.e., fifty percent of the customers cancelled their BT subscriptions while the other fifty percent remained with BT. For customers who cancelled their subscription the data represents the last journeys leading to cancellation within the time period of October to December 2019. To simplify our study, we chose a sample of 40 customers' journey data (20 from customers who cancelled and 20 from customers who did not), including only ``click"-related records, which totalled over 30,000 journeys in the sample data.

For each customer we had variable numbers of journeys (ranging from about 500 to over a thousand), and within each journey a variable number of events. The number of events within journeys varied from only one to over 500. The median number of events per journey was small, but the distribution was highly skewed to the right, with many journeys totalling more than 100 events. The total number of events in the sample data was approximately half a million.

The size of the dataset caused a problem with respect to memory allocation when fitting the model. Therefore, we proceeded with a subset of this data, using the first 300 journeys per customer, and up to the first 300 events within each journey, which amounted to a little over a third of the available data.

\subsection{Model Fitting and Analysis}

We model the mean number of events with a random intercept per customer, and include the random effects of the eight types of channels watched in a particular event within session in the linear predictor for $\eta_{isc}$. The linear predictors we use for the sample data described above are $$\log(\lambda_{sc} )=d_{0c}$$ and $$\eta_{isc}=\displaystyle\sum_{k=1}^Kb_{kc}\times\mbox{genre}_{k,isc},$$ where $\mbox{genre}_{k,isc}$ are dummy variables indicating whether event $i$ within session $s$ for customer $c$ was related to programme genre $k$. More specifically, there are eight major genres of IP channels on the BT TV box. However, to avoid intellectual property violations against BT or YouView, we denote them by numbers only ($G^{(1)}$ to $G^{(8)}$). The linear predictor used can be written as
\begin{align*}
\eta_{isc}&=b_{1c}\times G^{(1)}_{isc} + b_{2c}\times G^{(2)}_{isc} + b_{3c}\times G^{(3)}_{isc} + \\
& b_{4c}\times G^{(4)}_{isc} + b_{5c}\times G^{(5)}_{isc} + b_{6c}\times G^{(6)}_{isc} + \\
& b_{7c}\times G^{(7)}_{isc} + b_{8c}\times G^{(8)}_{isc},
\end{align*}
where $d_{0c}\sim\mbox{N}(\delta,\sigma_d^2)$, $b_{kc}\sim\mbox{N}(\beta_k,\sigma^2_{b_k})$, $k=1,\ldots,8$.

We estimate the model using the Bayesian framework, with 1,000 adaptation steps, 2,000 MCMC iterations as burn-in, and 20,000 MCMC iterations with a thinning of 10, for each of three MCMC chains. As prior distributions we use
\begin{align*}
\delta &\sim \mbox{N}(0, 1000) \\
\beta_k &\sim \mbox{N}(0, 1000) \\
\phi_c &\sim \mbox{N}(0, 1000) \\
\psi_c &\sim \mbox{Gamma}(0.001, 0.001) \\
\sigma_d^{-2} &\sim \mbox{Half-Cauchy} \\
\sigma_{b_k}^{-2} &\sim \mbox{Half-Cauchy} \\
\sigma_p^{-2} &\sim \mbox{Half-Cauchy} \\
\end{align*}
Model implementation was carried out using JAGS \cite{plummer} within R \cite{rcore}.

\subsection{Post-Processing and Benchmarking}

For each customer we compiled the estimates for the autocorrelation parameter $\phi$, the gamma distribution dispersion parameter $\psi$, the random intercepts $d_0$ and random genre effects $b_1$ to $b_8$. Therefore, each customer's profile can be described by these 11 variables. We then scaled each variable to have mean zero and unit variance. As exploratory analysis, we obtained a matrix of the Euclidean distances between customers and  performed hierarchical clustering using Ward's method, so as to minimise the intra-cluster variability. We then produced a dendrogram.

We used six different statistical machine learning methods to assess how accurate we could classify customers as ``active'' or ``cancelled'': (1) logistic regression (LR) \cite{nelder}, (2) random forests (RF) \cite{breiman}, (3) support vector machines (SVM) with a third-degree polynomial kernel \cite{svm}, (4) $\nu$-support vector classifiers ($\nu$-SVC) \cite{nusvc}, (5) $k$-nearest neighbours ($k$-NN) \cite{knn}, and (6) deep neural networks (DNN) \cite{dnn}. For the logistic regression we used a LASSO-type penalty \cite{lasso} to allow for mild regularisation of the predictor space, fixing the penalty parameter at $0.02$. For the random forests, we used 1,000 trees and a number of possible predictors to choose from equal to the floor of the square root of the total number of predictors. For KNN, we experimented with a different number of nearest neighbours and used $k=1$ in the end, which yielded the best performance. For DNN, after experimenting with different architectures, we used 17 hidden layers with 11, 12, 13, 14, 15, 14, 13, 12, 11, 10, 9, 8, 7, 6, 5, 4, and 3 neurons each, all using the rectified linear unit (ReLU) activation function; for the output layer we used the sigmoid activation function. For all methods, we used a 70:30 split for training and test data (therefore 28 customers for training the model and 12 for validation), and calculated the overall accuracy, and true and false positive rates as measures of predictive power of the machine learning methods.

We benchmarked the results from the machine learning methods against a na\"ive approach which consisted of using times between events within each journey for all customers as input for the machine learning algorithms described above. The times between events were used as predictors in the order journeys were recorded. Customers had an unequal number of recorded journeys, and journeys had unequal numbers of events. To circumvent the issue of incomplete cases, the minimum number of journeys per customer and minimum number of events per each journey were used. For example, if we had three customers with 20, 29 and 35 journeys each, we used data from the first 20 journeys for each customer. The same applies to events within journeys. The predictors were therefore the times between events in order of event occurrence within each journey, with journeys stacked as a vector. We refer to this approach as na\"ive for two reasons. First, customers could display different frustration signals, and recordings could have taken place at different stages of their frustration, and therefore the order with which journeys occur might be uninformative. Second, a large amount of information was lost due to restricting observations to generate complete cases, rather than describing each customer with the same sets of variables that reflect their overall profile (which is what the hierarchical Bayesian model proposed here attempts to do). In total, the benchmarking predictor space included 1,012 variables.

The implementation of the classification methods was made in Python using the libraries TensorFlow and Scikit-learn \cite{python}. The libraries NumPy and Pandas were also used for data manipulation and numerical analysis.

We provide code to implement the Bayesian model and machine learning methods at \url{https://github.com/rafamoral/profiling_tv_watching_behaviour}. However, we do not provide the raw data to avoid intellectual property violations against BT or YouView.

\section{Results}

\subsection{Customer Profiles}

The model took approximately 49 hours to run using four cores on an R Studio Server with an Intel Xeon 4614 2.2 GHz processor with 576 GB RAM, using three cores (one for each MCMC chain). Table~\ref{tab:estimates} displays the estimates for the $\delta$ and $\beta$ model parameters and variance components, associated standard errors and 95\% credible intervals, while Figure~\ref{fig:ranef} displays box plots for the customer-level parameter estimates ($\psi$ and $\phi$) and random effects.

\begin{table}[h]
    \centering
    \begin{tabular}{lrrr}
    \hline
    Parameter & \multicolumn{1}{c}{Estimate} & \multicolumn{2}{c}{95\% Credible interval} \\
    & \multicolumn{1}{c}{(std. error)} & \multicolumn{1}{c}{Lower bound} & \multicolumn{1}{c}{Upper bound} \\
    \hline
    $\delta$ & $2.49~(0.06)$ & $2.37$ & $2.61$ \\
    $\sigma_d$ & $0.39~(0.04)$ & $0.32$ & $0.49$ \\
    $\beta_1$ & $-0.19~(0.17)$ & $-0.53$ & $0.15$ \\
    $\beta_2$ & $-0.15~(0.19)$ & $-0.51$ & $0.22$ \\
    $\beta_3$ & $-0.24~(0.18)$ & $-0.60$ & $0.11$ \\
    $\beta_4$ & $-0.06~(0.17)$ & $-0.40$ & $0.28$ \\
    $\beta_5$ & $-0.48~(0.27)$ & $-0.99$ & $0.04$ \\
    $\beta_6$ & $-0.31~(0.21)$ & $-0.73$ & $0.11$ \\
    $\beta_7$ & $0.03~(0.20)$  & $-0.46$ & $0.41$ \\
    $\beta_8$ & $-0.94~(0.45)$ & $-1.80$ & $-0.01$ \\
    $\sigma_{b_1}$ & $1.10~(0.12)$ & $0.89$ & $1.37$ \\
    $\sigma_{b_2}$ & $1.15~(0.13)$ & $0.93$ & $1.44$ \\
    $\sigma_{b_3}$ & $1.11~(0.13)$ & $0.89$ & $1.39$ \\
    $\sigma_{b_4}$ & $1.06~(0.12)$ & $0.85$ & $1.33$ \\
    $\sigma_{b_5}$ & $1.50~(0.20)$ & $1.15$ & $1.93$ \\
    $\sigma_{b_6}$ & $1.26~(0.15)$ & $1.01$ & $1.60$ \\
    $\sigma_{b_7}$ & $1.13~(0.15)$ & $0.89$ & $1.45$ \\
    $\sigma_{b_8}$ & $0.98~(0.41)$ & $0.37$ & $1.95$ \\
    $\sigma_p$ & $0.23~(0.00)$ & $0.23$ & $0.24$ \\
    \hline
    \end{tabular}
    \caption{Parameter estimates, standard errors and 95\% credible intervals for the model fitted to the 40 customer sample data, using the first 300 journeys per customer and up to the first 300 events within each journey, which accounts for approximately a third of the data. We omit the estimates for $\phi_c$ and $\psi_c$, since there are 40 of each (one for each customer). They are shown, however, in Figure~\ref{fig:ranef}, alongside predicted random effects.}
    \label{tab:estimates}
\end{table}

\Figure[htb](topskip=0pt, botskip=0pt, midskip=0pt)[width = .48\textwidth]{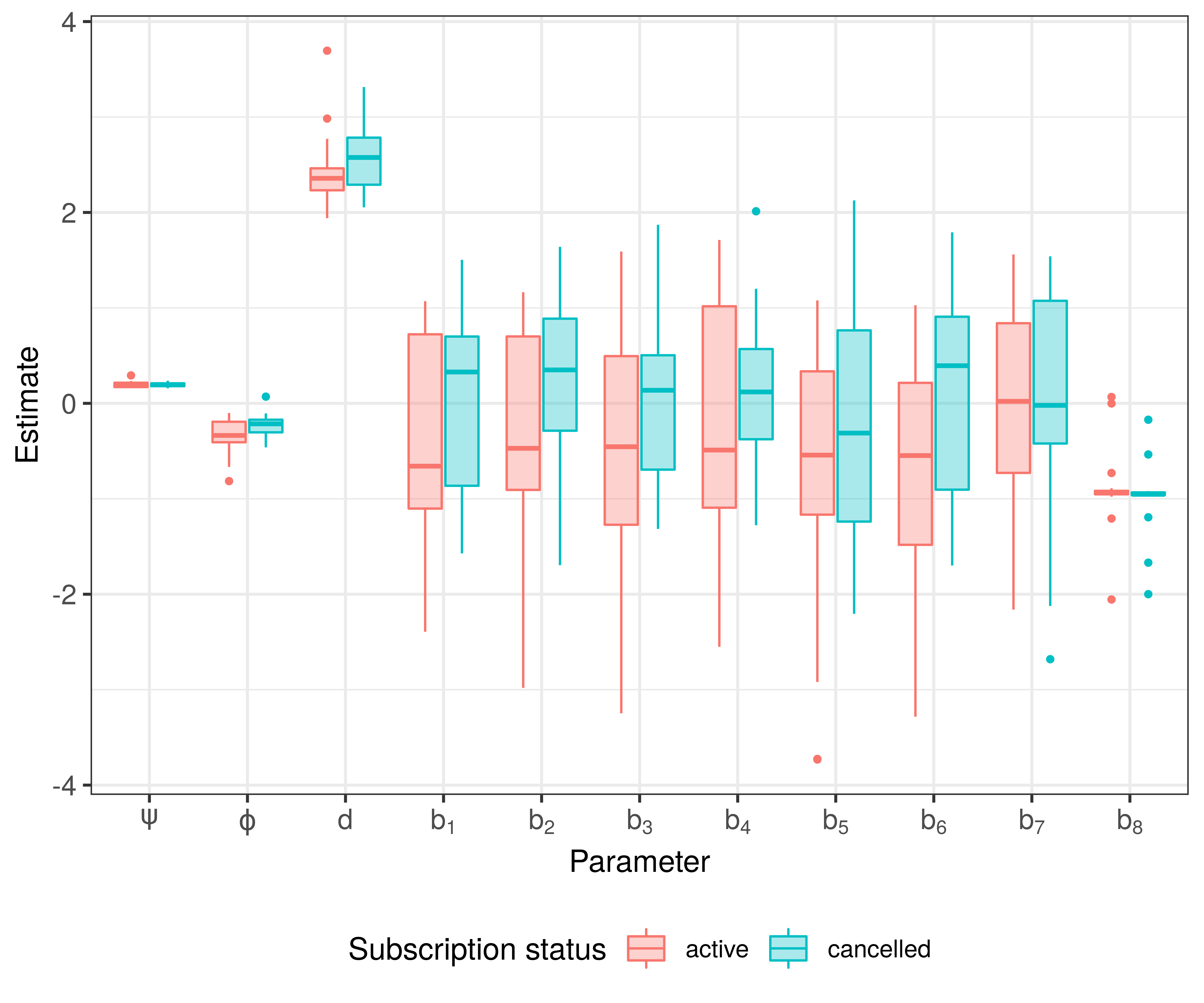}
{Box plots for the customer-level parameter estimates ($\psi$, the dispersion of time-between-events, and $\phi$, the mean autocorrelation between previous time-between-events and the time until the next event), and random effects ($b_1,\ldots,b_8$, corresponding to the effects of eight programme genres on time-between-events) for each of 40 customers, 20 having an active subscription to the TV service 20 users that cancelled the service.\label{fig:ranef}}

The means of the random effects corresponding to different programme genres are all close to zero, which means they play no significant role in predicting the time between events, apart from $\hat{\beta}_5$ (``Genre 5") and $\hat{\beta}_8$ (``Genre 8"), which indicate that the times between events are $1-e^{-0.48} = 38\%$ and $1-e^{-0.94} = 61\%$ shorter, on average, when these programme genres are being watched, respectively. For the other genres, the main predictor of time until the next event is the time between the two previous events, modelled by the autoregressive parameter. There is, however, a degree of individual variability, modelled by the variance components (which are close to 1 for all genres). Customers whose random effects are at the lower end could indicate a ``zapping" signature, which means they do not spend much time on a particular channel and instead, ``zap" to other channels, making the time between events short and the total number of events large. On the other hand, customers whose random effects are at the higher end would indicate more time spent watching one particular programme. This was seen more for the customers who cancelled their subscription (see Figure~\ref{fig:ranef}).

The dispersion parameter $\psi$ of the gamma distribution was small for all customers and did not vary much between them (Figure~\ref{fig:ranef}). Since the parameterisation of the gamma model used here implies that $\mbox{Var}(T) = \mu^2/\psi$, the model estimates indicate there was high variability in time between events. The variance components for the autoregressive parameter and number of events were small, indicating a smaller degree of individual variability between customers when looking at the dependence on time between previous events. However, on average the autoregressive random effects were slightly higher for customers who had cancelled their subscription (Figure~\ref{fig:ranef}).

The mean number of events $d$ was negatively correlated with all genre-related random effects $b$. This was expected, since a shorter time between events ultimately leads to more events taking place. The genre-related random effects $b$ were highly correlated between themselves, with this correlation being slightly weaker for customers in the ``cancelled" group. This suggests that television watching signatures did not change depending on programme genre. It is noteworthy to mention that the correlation between $b_8$ (``Genre 8") and all other genre-related random effects was much weaker, for customers both in the ``active" and ``cancelled" groups (see Figure~\ref{fig:corrplot}). The correlation is slightly smaller between $b_5$ (``Genre 5") and all other genre-related random effects for customers in the ``cancelled" group, which may help differentiate the two groups when applying machine learning methods.

\Figure[htb](topskip=0pt, botskip=0pt, midskip=0pt)[width = .48\textwidth]{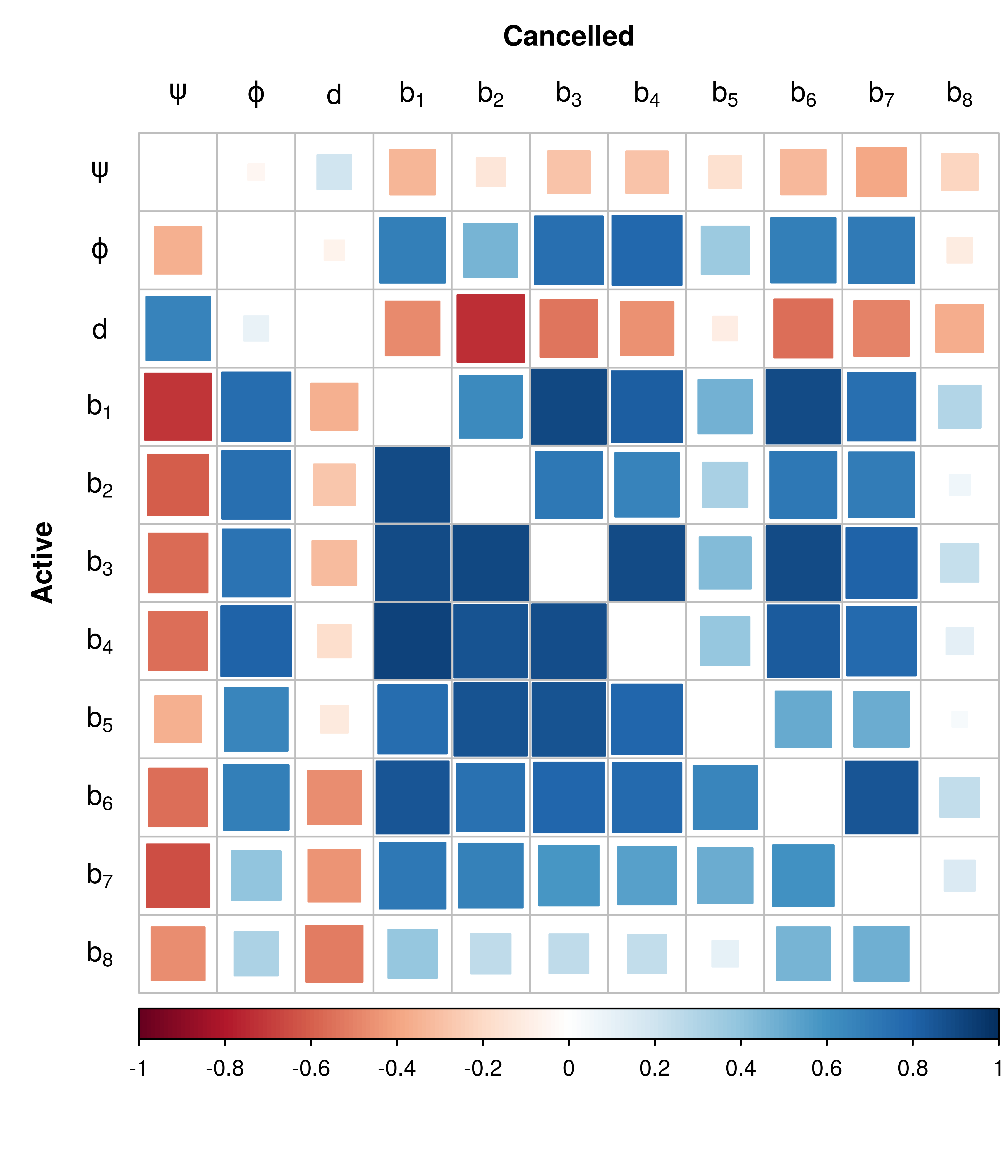}
{Correlation matrix of the estimated parameters and random effects per customer for the active group (lower triangular part) and group that cancelled the service (upper triangular part).\label{fig:corrplot}}

Correlations between the dispersion parameter estimates and other customer-level parameter estimates and random effects are all negative (apart from $d$), indicating that lower times between events yielded higher higher dispersion estimates, ultimately translating into a smaller variance (Figure~\ref{fig:corrplot}). Conversely, longer times between events yielded smaller dispersion estimates, which represent larger variances. This is straightforward to explain, for when ``zapping" variance will be smaller, since times are smaller and more consistent. If watching for longer times, they have scope to vary more. These correlations are much weaker for customers in the ``cancelled" group, showing that their ``zapping" behaviour was less intense. Overall variability remained unchanged when looking at customers that showed lower levels of activity in terms of triggering events versus those who spent more time watching offered content, and this can be helpful when attempting to identify an underlying frustration signature.

\subsection{Customer Clustering}

When clustering customers based on their raw times between events, the dendrogram does not show a clear separation between customers in the ``active" versus ``cancelled" groups. Even though Ward's method was used to minimise intra-cluster variability while maximising inter-cluster variability, the dendrogram seems to exhibit some level of chaining (see Figure~\ref{fig:tanglegram}, dendrogram on the left), and active customers are clustered alongside customers who cancelled their subscriptions at small Euclidean distances. The dendrogram produced using the customer-related estimated parameters and random effects from the Bayesian hierarchical model fit shows a much clearer two-cluster solution (see Figure~\ref{fig:tanglegram}, dendrogram on the right), and although there is no perfect separation between ``active" and ``cancelled" customer groups, at smaller Euclidean distances customers are generally clustered within their own groups.

\Figure[htb](topskip=0pt, botskip=0pt, midskip=0pt)[width = .9\textwidth]{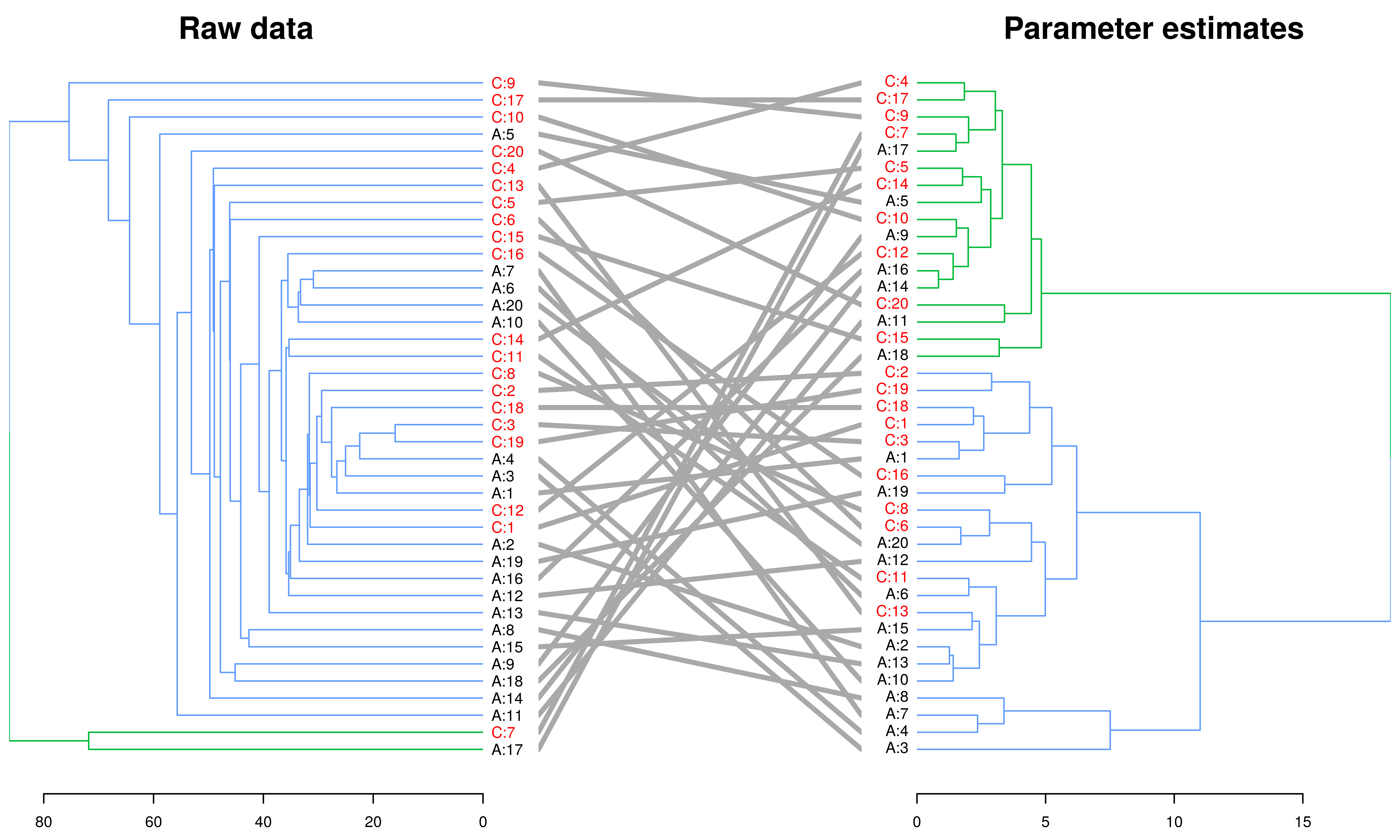}
{Dendrograms obtained from the hierarchical clustering applied to the raw data (left) and parameter estimates obtained from the modelling approach, indicating customer profiles (right). The x-axis represents the Euclidean distance between clusters. ``A" means that customer had an ``active" subscription and ``C" means the customer had ``cancelled" the service. The two-cluster solution is shown in different colours (green vs. blue), and the labels for customers in the ``cancelled" group are displayed in red for better visualisation. The gray lines connect the same customers to show their different positions in the two separate dendrograms.\label{fig:tanglegram}}

\subsection{Customer Classification}

When attempting to classify customer status (those who are ``active" vs. those who had ``cancelled" their subscription), the results obtained using the parameter estimates from the Bayesian hierarchical joint model were superior to the ones obtained using the raw data for all machine learning methods, in terms of accuracy, true and false positive rates (see Table~\ref{tab:class}). The SVM method yielded the best overall performance, with an accuracy of 92\% associated with 100\% true positive rate and 14\% false positive rate. Moreover, DNN and RF also presented 100\% true positive rates, however had a smaller accuracy due to mis-classification of customers in the ``active" group. Finally, $k$-NN was also associated with a low false positive rate, but mis-classified customers in the ``cancelled" group. The maximum accuracy obtained using the na\"ive approach was 58\%, using the $\nu$-SVC method; however, it was associated with a very low true positive rate (40\%).

\begin{table}[h]
    \centering
    \begin{tabular}{l|rrr|rrr}
    \hline
    \multirow{2}{*}{Method} & \multicolumn{3}{c|}{Model-based} & \multicolumn{3}{c}{Na\"ive} \\
    & Acc & TPR & FPR & Acc & TPR & FPR \\
    \hline
    SVM       & $0.92$ & $1.00$ & $0.14$ & $0.41$ & $1.00$ & $1.00$ \\
    DNN       & $0.83$ & $1.00$ & $0.29$ & $0.42$ & $1.00$ & $1.00$ \\
    $\nu$-SVC & $0.75$ & $0.80$ & $0.29$ & $0.58$ & $0.40$ & $0.29$ \\
    $k$-NN    & $0.75$ & $0.60$ & $0.14$ & $0.33$ & $0.80$ & $1.00$ \\
    RF        & $0.67$ & $1.00$ & $0.57$ & $0.25$ & $0.40$ & $0.86$ \\
    LASSO-LR  & $0.67$ & $0.80$ & $0.57$ & $0.25$ & $0.49$ & $0.86$ \\
    \hline
    \end{tabular}
    \caption{Accuracy (Acc), true positive rates (TPR) and false positive rates (FPR) for each of six machine learning methods applied to the classification of customer status (``active" vs. ``cancelled") based on the parameter estimates obtained from the Bayesian hierarchical modelling approach (model-based) or the raw data (na\"ive). We used a 70:30 split between training:validation sets, i.e. the training set contained 28 customers and the validation set contained 12 customers. All measures reported on this table were obtained in the validation set. SVM: support vector machines with a third-degree polynomial kernel; DNN: deep neural networks; $\nu$-SVC: $\nu$-support vector classifier; $k$-NN: $k$-nearest neighbours; RF: random forests; LASSO-LR: logistic regression with LASSO-type penalty.}
    \label{tab:class}
\end{table}

\section{Discussion}

We proposed a novel Bayesian hierarchical joint model to analyse customer journeys. This model is able to characterise customer profiles based on how many events take place within different journeys and also the time taken between events. The model drastically reduces the dimensionality of the data from thousands of observations per customer to 11 customer-level parameter estimates and random effects. This allows for prompt application of many different machine learning methods for classification, ultimately allowing for churn prediction.

Our results show that our model reflects the underlying processes well, and is a sensible representation of a customer's profile. Therefore it is able to retain the most prominent features of the data, and is a better candidate for machine learning techniques rather than using the high-dimensional raw data. Multicollinearity between genre-related random effects may represent an additional challenge when employing machine learning methods, however that was circumvented by the classification tools used in different ways (e.g., regularisation when using LASSO-type penalties, random selection of predictors within a random forests framework, incorporation of non-linearities when splitting the parametric space within a support vector classifier framework, etc).

The strengths of our proposed methodology are evidenced by the high classification accuracy obtained. Obtaining customer profiles based on the Bayesian hierarchical joint model was therefore a successful way of dealing with the high-dimensional journey data. Not only does this represent promising results from a classification perspective, it also provides interpretable machine learning aimed at characterising customer churn. For instance, from a random forests model, we may look at different measures of variable importance, whereas in a logistic regression framework a LASSO penalty will automatically select variables that are most important. Moreover, the estimates from the Bayesian model themselves are characterisations of each customer's profile, as well as how different covariates may affect the behaviour of the time between events and number of events in their journeys. This makes the novel model proposed here useful not only for post-processing aimed at churn prediction, but also for using customer status as a predictor variable and studying, in hindsight, what are the most prominent features that may have led to churn.

We identify two main limitations to this study. First, the number of customers available in the dataset is small. A larger pool of customers would allow for better validation of the methods. However, this does serve as proof of concept and already shows great promise, for with a small sample the methodology showed a high predictive power. Second, fitting the Bayesian hierarchical joint model takes a high toll in terms of computational burden. Given the high-dimensionality of the data, allocating the necessary memory proves to be a difficult task, even when using a high-spec server. Nevertheless, even when using about a third of the data, we already obtained high predictive power, again demonstrating that our proposed framework is capable of accurately describing customer profiles.

Future work involves (1) overcoming memory limitations to run the model for a larger sample of customers, including more journeys and events; and (2) carrying out simulation studies to assess the trade-offs between using more journeys versus more events within journeys. By understanding how these trade-offs work, we may use fewer journeys and/or events, therefore being able to analyse a larger pool of customers without losing much in terms of descriptive and predictive power.

\EOD

\end{document}